\documentclass[11pt]{article}
\usepackage{amssymb,amsmath,amsfonts}
\usepackage{graphicx}
\usepackage{graphics}
\usepackage{eepic,epsfig}

\begin{document}

\title{Casimir densities in brane models with compact internal spaces}
\author{A. A. Saharian\thanks{%
Email: saharian@ictp.it} \\
\textit{Department of Physics, Yerevan State University,} \\
\textit{1 Alex Manoogian Street, 0025 Yerevan, Armenia}}
\date{\today}
\maketitle

\begin{abstract}
We investigate the Wightman function, the vacuum expectation values of the
field squared and the energy-momentum tensor for a massive scalar field with
general curvature coupling parameter subject to Robin boundary conditions on
two codimension one parallel branes located on $(D+1)$-dimensional
background spacetime $AdS_{D_{1}+1}\times \Sigma $ with a warped internal
space $\Sigma $. The general case of different Robin coefficients on
separate branes is considered. Unlike to the purely AdS bulk, the vacuum
expectation values induced by a single brane, in addition to the distance
from the brane, depends also on the position of the brane in the bulk. The
brane induced parts in these expectation values vanish when the brane
position tends to the AdS horizon or AdS boundary. For strong gravitational
fields corresponding to large values of the AdS energy scale, the both
single brane and interference parts of the expectation values integrated
over the internal space are exponentially suppressed. An application to the
higher dimensional generalization of the Randall-Sundrum brane model with
arbitrary mass terms on the branes is discussed. For large distances between
the branes the induced surface densities give rise to an exponentially
suppressed cosmological constant on the brane.
\end{abstract}

\bigskip


\bigskip

\section{Introduction}

\label{sec:introd}

The braneworld scenario provides an interesting alternative to the standard
Kaluza-Klein compactification of the extra dimensions. The simplest
phenomenological models describing such a scenario are the five-dimensional
Randall-Sundrum type braneworld models (for a review see \cite{Ruba01}).
From the point of view of embedding these models into a more fundamental
theory, such as string/M-theory, one may expect that a more complete version
of the scenario must admit the presence of additional extra dimensions
compactified on an internal manifold. From a phenomenological point of view,
the consideration of more general spacetimes offer a richer geometrical
structure and may provide interesting extensions of the Randall-Sundrum
mechanism for the geometric origin of the hierarchy. More extra dimensions
also relax the fine-tunings of the fundamental parameters. These models can
provide a framework in the context of which the stabilization of the radion
field naturally takes place. In addition, a richer topological structure of
the field configuration in transverse space provides the possibility of more
realistic spectrum of chiral fermions localized on the brane. Several
variants of the Randall-Sundrum scenario involving cosmic strings and other
global defects of various codimensions have been investigated in higher
dimensions (see, for instance, \cite{Greg00} and references therein).

Motivated by the problems of the radion stabilization and the
generation of cosmological constant, the role of quantum effects
in braneworlds has attracted great deal of attention
\cite{Fabi00}-\cite{Mina07}. A class of higher dimensional models
with the topology $\mathrm{AdS}_{D_{1}+1}\times \Sigma $, where
$\Sigma $ is a one-parameter compact manifold, and with two branes
of codimension one located at the orbifold fixed points, is
considered in Refs. \cite{Flac03b,Flac03}. In both cases of the
warped and unwarped internal manifold, the quantum effective
potential induced by bulk scalar fields is evaluated and it has
been shown that this potential can stabilize the hierarchy between
the Planck and electroweak scales without fine tuning. In addition
to the effective potential, the investigation of local physical
characteristics in these models is of considerable interest. Local
quantities contain more information on the vacuum fluctuations
than the global ones and play an important role in modelling a
self-consistent
dynamics involving the gravitational field. In papers \cite%
{Saha06a,Saha06b,Saha07} we have studied the bulk and surface Casimir
densities for a scalar field with an arbitrary curvature coupling parameter
obeying Robin boundary conditions on two codimension one parallel branes
embedded in the background spacetime $\mathrm{AdS}_{D_{1}+1}\times \Sigma $
with a warped internal space $\Sigma $. For an arbitrary internal space $%
\Sigma $, the application of the generalized Abel-Plana formula \cite%
{SahaRev} allowed us to extract form the vacuum expectation values the part
due to the bulk without branes and to present the brane induced parts in
terms of exponentially convergent integrals for the points away from the
branes. In the present paper we review these results.

The paper is organized as follows. In the next section we evaluate the
Wightman function in the region between the branes. By using the generalized
Abel-Plana formula, we present this function in the form of a sum of the
Wightman function for the bulk without boundaries and boundary induced
parts. The vacuum expectation value of the bulk energy-momentum tensor for a
general case of the internal space $\Sigma $ is discussed in section \ref%
{sec:EMT}. The interaction forces between the branes are discussed in
section \ref{sec:Intforce}. The surface Casimir densities and the energy
balance are considered in section \ref{sec:SurfEMT}. The last section
contains a summary of the work.

\section{Wightman function}

\label{sec:WF}

For a free scalar field $\varphi (x)$ with curvature coupling parameter $%
\zeta $ the equation of motion has the form
\begin{equation}
\left( g^{MN}\nabla _{M}\nabla _{N}+m^{2}+\zeta R\right) \varphi (x)=0,
\label{fieldeq}
\end{equation}%
where $M,N=0,1,\ldots ,D$, and $R$ is the scalar curvature. We will assume
that the background spacetime has a topology $AdS_{D_{1}+1}\times \Sigma $,
where $\Sigma $ is a $D_{2}$-dimensional compact manifold. The corresponding
line element has the form
\begin{equation}
ds^{2}=g_{MN}dx^{M}dx^{N}=e^{-2k_{D}y}\eta _{\mu \nu }dx^{\mu }dx^{\nu
}-e^{-2k_{D}y}\gamma _{ij}dX^{i}dX^{j}-dy^{2},  \label{metric}
\end{equation}%
with $\eta _{\mu \nu }=\mathrm{diag}(1,-1,\ldots ,-1)$ being the metric for
the $D_{1}$-dimensional Minkowski spacetime $R^{(D_{1}-1,1)}$ and the
coordinates $X^{i}$ cover the manifold $\Sigma $, $D=D_{1}+D_{2}$. Here and
below $\mu ,\nu =0,1,\ldots ,D_{1}-1$ and $i,j=1,\ldots ,D_{2}$. The scalar
curvature for the metric tensor from (\ref{metric}) is given by the
expression $R=-D(D+1)k_{D}^{2}-e^{2k_{D}y}R_{(\gamma )}$, where $R_{(\gamma
)}$ is the scalar curvature for the metric tensor $\gamma _{ik}$. In the
discussion below, in addition to the coordinate $y$\ we will use the radial
coordinate $z$ defined by the relation $z=e^{k_{D}y}/k_{D}$. In terms of the
coordinate $z$, the metric tensor is conformally related to the metric of
the direct product space $R^{(D_{1},1)}\times \Sigma $ by the conformal
factor $(k_{D}z)^{-2}$.

Our main interest in this paper will be the Wightman function and the vacuum
expectation values (VEVs) of the field squared and the energy-momentum
tensor induced by two infinite parallel branes of codimension one with the
coordinates $y=a$ and $y=b$, $a<b$. We will assume that on this branes the
scalar field obeys the boundary conditions
\begin{equation}
(\tilde{A}_{j}+\tilde{B}_{j}\partial _{y})\varphi (x)=0,\quad y=j,\;j=a,b,
\label{boundcond}
\end{equation}%
with constant coefficients $\tilde{A}_{j}$, $\tilde{B}_{j}$. In the
orbifolded version of the model which corresponds to a higher dimensional
Randall-Sundrum braneworld these coefficients are expressed in terms of the
surface mass parameters and the curvature coupling of the scalar field. In
quantum field theory the imposition of boundary conditions modifies the
spectrum for the zero--point fluctuations and as a result the VEVs for
physical observables are changed. These effects can either stabilize or
destabilize the branewolds and have to be taken into account in the
self-consistent formulation of the braneworld dynamics.

As a first stage in the investigations of local quantum effects, we will
consider the positive frequency Wightman function defined as the expectation
value $G^{+}(x,x^{\prime })=\langle 0|\varphi (x)\varphi (x^{\prime
})|0\rangle $. In the region between the branes, $a<y<b$, the Wightman
function is presented as the mode-sum:%
\begin{eqnarray}
G^{+}(x,x^{\prime }) &=&\frac{k_{D}^{D-1}(zz^{\prime })^{D/2}}{%
2^{D_{1}+1}\pi ^{D_{1}-3}z_{a}^{2}}\sum_{\beta }\psi _{\beta }(X)\psi
_{\beta }^{\ast }(X^{\prime })\int d\mathbf{k}\,e^{i\mathbf{k}\Delta \mathbf{%
x}}  \notag \\
&\times &\sum_{n=1}^{\infty }\frac{h_{\beta \nu }(u)}{\left[
A_{b}^{2}+B_{b}^{2}(\eta ^{2}u^{2}-\nu ^{2})\right] \bar{J}_{\nu }^{(a)2}(u)/%
\bar{J}_{\nu }^{(b)2}(\eta u)-A_{a}^{2}+B_{a}^{2}(u^{2}-\nu ^{2})}%
|_{u=\gamma _{\nu ,n}},  \label{W11}
\end{eqnarray}%
where $\mathbf{x}=(x^{1},x^{2},\ldots ,x^{D_{1}-1})$ represents the spatial
coordinates in $R^{(D_{1}-1,1)}$, $\Delta \mathbf{x}=\mathbf{x}-\mathbf{x}%
^{\prime }$, $\eta =z_{b}/z_{a}$, and%
\begin{eqnarray}
h_{\beta \nu }(u) &=&ug_{\nu }(u,uz/z_{a})g_{\nu }(u,uz^{\prime }/z_{a})%
\frac{e^{-i\Delta t\sqrt{u^{2}/z_{a}^{2}+k^{2}+\lambda _{\beta }^{2}}}}{%
\sqrt{u^{2}/z_{a}^{2}+k^{2}+\lambda _{\beta }^{2}}},  \label{hab} \\
g_{\nu }(u,v) &=&J_{\nu }(v)\bar{Y}_{\nu }^{(a)}(u)-\bar{J}_{\nu
}^{(a)}(u)Y_{\nu }(v),\;\nu =\sqrt{(D/2)^{2}-D(D+1)\zeta +m^{2}/k_{D}^{2}},
\label{genu}
\end{eqnarray}%
with $k=|\mathbf{k}|$, $\Delta t=t-t^{\prime }$, $z_{j}=e^{k_{D}j}/k_{D}$, $%
j=a,b$, $J_{\nu }(x)$, $Y_{\nu }(x)$ are the Bessel and Neumann functions.
In formula (\ref{genu}), for a given function $F(x)$ we use the notation
\begin{equation}
\bar{F}^{(j)}(x)=A_{j}F(x)+B_{j}xF^{\prime }(x),\;A_{j}=\tilde{A}_{j}+\tilde{%
B}_{j}k_{D}D/2,\quad B_{j}=\tilde{B}_{j}k_{D},\;j=a,b.  \label{notbar}
\end{equation}%
In the discussion below we will assume values of the curvature coupling
parameter for which $\nu $ is real. For imaginary $\nu $ the ground state
becomes unstable \cite{Brei82}. In (\ref{W11}), the modes $\psi _{\beta }(X)$
are the eigenfunctions for the operator $\Delta _{(\gamma )}+\zeta
R_{(\gamma )}$:
\begin{equation}
\left[ \Delta _{(\gamma )}+\zeta R_{(\gamma )}\right] \psi _{\beta
}(X)=-\lambda _{\beta }^{2}\psi _{\beta }(X),\;\int d^{D_{2}}X\,\sqrt{\gamma
}\psi _{\beta }(X)\psi _{\beta ^{\prime }}^{\ast }(X)=\delta _{\beta \beta
^{\prime }},  \label{eqint1}
\end{equation}%
with eigenvalues $\lambda _{\beta }^{2}$, and $\Delta _{(\gamma )}$ is the
Laplace-Beltrami operator for the metric $\gamma _{ij}$. From the boundary
condition on the branes we receive that the eigenvalues $\gamma _{\nu ,n}$
have to be solutions to the equation
\begin{equation}
g_{\nu }^{(ab)}(\gamma _{\nu ,n},\eta \gamma _{\nu ,n})\equiv \bar{J}_{\nu
}^{(a)}(\gamma _{\nu ,n})\bar{Y}_{\nu }^{(b)}(\eta \gamma _{\nu ,n})-\bar{Y}%
_{\nu }^{(a)}(\gamma _{\nu ,n})\bar{J}_{\nu }^{(b)}(\eta \gamma _{\nu ,n})=0.
\label{cnu}
\end{equation}%
This equation determines the tower of radial Kaluza-Klein (KK) masses.

Applying to the sum over $n$ in (\ref{W11}) a variant of the generalized
Abel-Plana formula \cite{SahaRev}, the Wightman function is presented in two
equivalent forms ($j=a,b$)
\begin{eqnarray}
G^{+}(x,x^{\prime }) &=&G_{0}^{+}(x,x^{\prime })+\langle \varphi (x)\varphi
(x^{\prime })\rangle ^{(j)}-\frac{k_{D}^{D-1}(zz^{\prime })^{D/2}}{%
2^{D_{1}-1}\pi ^{D_{1}}}\sum_{\beta }\psi _{\beta }(X)\psi _{\beta }^{\ast
}(X^{\prime })  \notag \\
&\times &\int d\mathbf{k}\,e^{i\mathbf{k}\Delta \mathbf{x}}\int_{\sqrt{%
k^{2}+\lambda _{\beta }^{2}}}^{\infty }duuG_{\nu }^{(j)}(uz_{a},uz)G_{\nu
}^{(j)}(uz_{a},uz^{\prime })  \notag \\
&\times &\frac{\Omega _{j\nu }(uz_{a},uz_{b})}{\sqrt{u^{2}-k^{2}-\lambda
_{\beta }^{2}}}\cosh (\Delta t\sqrt{u^{2}-k^{2}-\lambda _{\beta }^{2}}),
\label{W15}
\end{eqnarray}%
where $I_{\nu }(u)$ and $K_{\nu }(u)$ are the modified Bessel functions and
\begin{eqnarray}
\Omega _{a\nu }(u,v) &=&\frac{\bar{K}_{\nu }^{(b)}(v)/\bar{K}_{\nu }^{(a)}(u)%
}{\bar{K}_{\nu }^{(a)}(u)\bar{I}_{\nu }^{(b)}(v)-\bar{K}_{\nu }^{(b)}(v)\bar{%
I}_{\nu }^{(a)}(u)},  \notag \\
\Omega _{b\nu }(u,v) &=&\frac{\bar{I}_{\nu }^{(a)}(u)/\bar{I}_{\nu }^{(b)}(v)%
}{\bar{K}_{\nu }^{(a)}(u)\bar{I}_{\nu }^{(b)}(v)-\bar{K}_{\nu }^{(b)}(v)\bar{%
I}_{\nu }^{(a)}(u)},  \label{Omnub} \\
G_{\nu }^{(j)}(u,v) &=&I_{\nu }(v)\bar{K}_{\nu }^{(j)}(u)-\bar{I}_{\nu
}^{(j)}(u)K_{\nu }(v),\;j=a,b.  \notag
\end{eqnarray}%
In (\ref{W15}), the term
\begin{eqnarray}
G_{0}^{+}(x,x^{\prime }) &=&\frac{k_{D}^{D-1}(zz^{\prime })^{\frac{D}{2}}}{%
2^{D_{1}}\pi ^{D_{1}-1}}\sum_{\beta }\psi _{\beta }(X)\psi _{\beta }^{\ast
}(X^{\prime })\int d\mathbf{k}\,e^{i\mathbf{k}\Delta \mathbf{x}}  \notag \\
&\times &\int_{0}^{\infty }du\,u\frac{e^{-i\Delta t\sqrt{u^{2}+k^{2}+\lambda
_{\beta }^{2}}}}{\sqrt{u^{2}+k^{2}+\lambda _{\beta }^{2}}}J_{\nu }(uz)J_{\nu
}(uz^{\prime }),  \label{WAdS}
\end{eqnarray}%
does not depend on the boundary conditions and is the Wightman function for
the $AdS_{D_{1}+1}\times \Sigma $ spacetime without branes. The second term
on the right of Eq. (\ref{W15}) is given by the formula
\begin{eqnarray}
\langle \varphi (x)\varphi (x^{\prime })\rangle ^{(a)} &=&-\frac{%
k_{D}^{D-1}(zz^{\prime })^{\frac{D}{2}}}{2^{D_{1}-1}\pi ^{D_{1}}}\sum_{\beta
}\psi _{\beta }(X)\psi _{\beta }^{\ast }(X^{\prime })\int d\mathbf{k}\,e^{i%
\mathbf{k}\Delta \mathbf{x}}  \notag \\
&\times &\int_{\sqrt{k^{2}+\lambda _{\beta }^{2}}}^{\infty }duu\frac{\bar{I}%
_{\nu }^{(a)}(uz_{a})}{\bar{K}_{\nu }^{(a)}(uz_{a})}\frac{K_{\nu }(uz)K_{\nu
}(uz^{\prime })}{\sqrt{u^{2}-k^{2}-\lambda _{\beta }^{2}}}\cosh \!(\Delta t%
\sqrt{u^{2}-k^{2}-\lambda _{\beta }^{2}}),  \label{1bounda}
\end{eqnarray}%
for $j=a$, and the expression for $\langle \varphi (x)\varphi (x^{\prime
})\rangle ^{(b)}$ is obtained from (\ref{1bounda}) by the replacements $%
a\rightarrow b$, $I_{\nu }\rightleftarrows K_{\nu }$. The term $\langle
\varphi (x)\varphi (x^{\prime })\rangle ^{(j)}$\ does not depend on the
parameters of the brane at $z=z_{j^{\prime }}$, $j^{\prime }\neq j$, and is
induced by a single brane at $z=z_{j}$ when the boundary $z=z_{j^{\prime }}$
is absent. In the same way described above for the Wightman function, any
other two-point function can be evaluated. Note that the expression for the
Wightman function is not symmetric with respect to the interchange of the
brane indices. The reason for this is that the boundaries have nonzero
extrinsic curvature tensors and two sides of the boundaries are not
equivalent. In particular, for the geometry of a single brane the VEVs are
different for the regions on the left and on the right of the brane. In the
region $y<a$ the Wightman has the form $G^{+}(x,x^{\prime
})=G_{0}^{+}(x,x^{\prime })+\langle \varphi (x)\varphi (x^{\prime })\rangle
^{(a)}$, where the expression for the second term on the right hand-side is
obtained from (\ref{1bounda}) by the replacement $I_{\nu }\rightleftarrows
K_{\nu }$. Similarly, for the Wightman function in the region $y>b$ one has $%
G^{+}(x,x^{\prime })=G_{0}^{+}(x,x^{\prime })+\langle \varphi (x)\varphi
(x^{\prime })\rangle ^{(b)}$, where the second term is given by formula (\ref%
{1bounda}) replacing $a\rightarrow b$.

In the higher dimensional generalization of the Randall-Sundrum braneworld
based on the bulk $AdS_{D_{1}+1}\times \Sigma $ the Wightman function for
untwisted scalar is given by formula (\ref{W15}) with an additional factor
1/2 and with Robin coefficients
\begin{equation}
\tilde{A}_{a}/\tilde{B}_{a}=-c_{a}/2-2D\zeta k_{D},\quad \tilde{A}_{b}/%
\tilde{B}_{b}=c_{b}/2-2D\zeta k_{D}2.  \label{AtildeRS}
\end{equation}%
For twisted scalar field Dirichlet boundary conditions are obtained. The
one-loop effective potential and the problem of moduli stabilization in this
model with zero mass parameters $c_{j}$ are discussed in Ref. \cite{Flac03b}.

\section{Vacuum energy-momentum tensor}

\label{sec:EMT}

The VEV of the energy-momentum tensor can be evaluated by substituting the
Wightman function and the VEV of the field squared into the formula
\begin{equation}
\langle 0|T_{MN}|0\rangle =\lim_{x^{\prime }\rightarrow x}\partial
_{M}\partial _{N}^{\prime }G^{+}(x,x^{\prime })+\left[ \left( \zeta -\frac{1%
}{4}\right) g_{MN}\nabla _{L}\nabla ^{L}-\zeta \nabla _{M}\nabla _{N}-\zeta
R_{MN}\right] \langle 0|\varphi ^{2}|0\rangle ,  \label{vevEMT1pl}
\end{equation}%
where $R_{MN}$ is the Ricci tensor. Substituting the expression for the
Wightman function into this formula, for the components of the vacuum
energy-momentum tensor in the region between the branes we obtain the
formula
\begin{eqnarray}
\langle 0|T_{M}^{N}|0\rangle &=&\langle T_{M}^{N}\rangle ^{(0)}+\langle
T_{M}^{N}\rangle ^{(j)}-\frac{2k_{D}^{D+1}z^{D}}{(4\pi )^{D_{1}/2}\Gamma
\left( D_{1}/2\right) }\sum_{\beta }|\psi _{\beta }(X)|^{2}  \notag \\
&&\times \int_{\lambda _{\beta }}^{\infty }du\,u(u^{2}-\lambda _{\beta
}^{2})^{\frac{D_{1}}{2}-1}\Omega _{j\nu }(uz_{a},uz_{b})F_{\beta
M}^{(+)N}[G_{\nu }^{(j)}(uz_{j},uz)],  \label{Tikjint}
\end{eqnarray}%
with the functions $F_{\beta M}^{(+)N}[g(v)]$, $g(v)=G_{\nu
}^{(j)}(uz_{j},v) $, defined by the relations
\begin{eqnarray}
F_{\beta \mu }^{(\pm )\sigma }[g(v)] &=&\delta _{\mu }^{\sigma }\left( \frac{%
1}{4}-\zeta \right) \left\{ z^{2}g^{2}(v)\eta _{\beta }(X)+2v\frac{\partial
}{\partial v}F[g(v)]+\frac{\pm v^{2}-z^{2}\lambda _{\beta }^{2}}{D_{1}(\zeta
-1/4)}g^{2}(v)\right\} ,  \label{Fmu} \\
F_{\beta D}^{(\pm )D}[g(v)] &=&\left( \frac{1}{4}-\zeta \right)
z^{2}g^{2}(v)\eta _{\beta }(X)+\frac{1}{2}[-v^{2}g^{\prime 2}(v)  \notag \\
&&+D(4\zeta -1)vg(v)g^{\prime }(v)+\left( 2m^{2}/k_{D}^{2}-\nu ^{2}\pm
v^{2}\right) g^{2}(v)],  \label{FD}
\end{eqnarray}%
for the components in the AdS part, and by the relations
\begin{eqnarray}
F_{\beta D}^{(\pm )i}[g(v)] &=&\frac{k_{D}}{2}z^{2}(1-4\zeta )F[g(v)]\eta
_{\beta }^{i}(X),  \label{FiD} \\
F_{\beta i}^{(\pm )k}[g(v)] &=&z^{2}g^{2}(v)\frac{t_{\beta i}^{k}(X)}{|\psi
_{\beta }(X)|^{2}}+\frac{1}{2}\delta _{i}^{k}(1-4\zeta )v\frac{\partial }{%
\partial v}F[g(v)],  \label{Fi}
\end{eqnarray}%
with $t_{\beta i}^{k}(X)=-\gamma ^{kl}t_{\beta il}(X)$, for the components
having indices in the internal space. In these expressions we use the
following notations
\begin{eqnarray}
F[g(v)] &=&vg(v)g^{\prime }(v)+\frac{1}{2}\left( D+\frac{4\zeta }{4\zeta -1}%
\right) g^{2}(v),  \label{Fgv} \\
\eta _{\beta }(X) &=&\frac{\triangle _{(\gamma )}|\psi _{\beta }(X)|^{2}}{%
|\psi _{\beta }(X)|^{2}},\quad \eta _{\beta }^{i}(X)=-\gamma ^{ik}\frac{%
\partial _{k}|\psi _{\beta }(X)|^{2}}{|\psi _{\beta }(X)|^{2}},  \label{etaX}
\\
t_{\beta ik}(X) &=&\nabla _{(\gamma )i}\psi _{\beta }(X)\nabla _{(\gamma
)k}\psi _{\beta }^{\ast }(X)+  \notag \\
&&\left[ \left( \zeta -\frac{1}{4}\right) \gamma _{ik}\triangle _{(\gamma
)}-\zeta \nabla _{(\gamma )i}\nabla _{(\gamma )k}-\zeta R_{(\gamma )ik}%
\right] |\psi _{\beta }(X)|^{2},  \label{tbetik}
\end{eqnarray}%
where $\nabla _{(\gamma )i}$ is the covariant derivative operator associated
with the metric tensor $\gamma _{ik}$.

In formula (\ref{Tikjint}),
\begin{equation}
\langle T_{M}^{N}\rangle ^{(0)}=\frac{k_{D}^{D+1}z^{D}}{(4\pi )^{\frac{D_{1}%
}{2}}}\Gamma \left( 1-\frac{D_{1}}{2}\right) \sum_{\beta }|\psi _{\beta
}(X)|^{2}\int_{0}^{\infty }du\,u(u^{2}+\lambda _{\beta }^{2})^{\frac{D_{1}}{2%
}-1}F_{\beta M}^{(-)N}[J_{\nu }(uz)],  \label{EMTAdS}
\end{equation}%
is the VEV for the energy-momentum tensor in the background without branes,
and the term $\langle T_{M}^{N}\rangle ^{(j)}$ is induced by a single brane
at $z=z_{j}$. For the left brane one has
\begin{equation}
\langle T_{M}^{N}\rangle ^{(a)}=-\frac{2k_{D}^{D+1}z^{D}}{(4\pi )^{\frac{%
D_{1}}{2}}\Gamma \left( \frac{D_{1}}{2}\right) }\sum_{\beta }|\psi _{\beta
}(X)|^{2}\int_{\lambda _{\beta }}^{\infty }du\,u(u^{2}-\lambda _{\beta
}^{2})^{\frac{D_{1}}{2}-1}\frac{\bar{I}_{\nu }^{(a)}(uz_{a})}{\bar{K}_{\nu
}^{(a)}(uz_{a})}F_{\beta M}^{(+)N}[K_{\nu }(uz)],  \label{EMT1bounda}
\end{equation}%
and the corresponding expression for the right brane is obtained by the
replacements $a\rightarrow b$, $I_{\nu }\rightleftarrows K_{\nu }$. Unlike
to the case of purely AdS bulk, here the VEVs for a single brane in addition
to the distance from the brane depend also on the position of the brane in
the bulk. In the limit when the AdS curvature radius tends to infinity we
derive the formula for the vacuum energy-momentum tensor for parallel plates
on the background spacetime with topology $R^{(D_{1},1)}\times \Sigma $. In
this limit for a homogeneous internal space ${}_{D}^{D}$--component of the
brane induced part in the VEV of the energy-momentum tensor vanishes.

The features of the single brane parts in the VEVs in the asymptotic regions
of the parameters are as follows. For the points on the brane the vacuum
energy-momentum tensor diverges. Near the brane the total vacuum
energy-momentum tensor is dominated by the brane induced part and has
opposite signs for Dirichlet and non-Dirichlet boundary conditions. Near the
brane ${}_{D}^{D}$-- and ${}_{D}^{i}$--components of this tensor have
opposite signs in the regions $y<a$ and $y>a$. For large distances from the
brane in the region $y>a$ the contribution of a given mode along $\Sigma $
with nonzero KK mass is suppressed by the factor $e^{-2\lambda _{\beta }z}$.
For the zero mode the brane induced VEV near the AdS horizon behaves as $%
z^{D_{2}-2\nu }$. In the purely AdS bulk ($D_{2}=0$) this VEV vanishes on
the horizon for $\nu >0$. For an internal spaces with $D_{2}>2\nu $ the VEV
diverges on the horizon. The VEV integrated over the internal space vanishes
on the AdS horizon for all values $D_{2}$ due to the additional warp factor
coming from the volume element. For the points near the AdS boundary, the
brane induced VEV vanishes as $z^{D+2\nu }$ for diagonal components and as $%
z^{D+2\nu +2}$ for the ${}_{D}^{i}$--component. For small values of the
length scale for the internal space, the contribution of nonzero KK masses
is exponentially suppressed and the main contribution into the brane induced
energy-momentum tensor comes from the zero mode. In the opposite limit, when
the length scale of the internal space is large, to the leading order the
vacuum energy-momentum tensor reduces to the corresponding result for a
brane in the bulk $AdS_{D+1}$ given in Ref. \cite{Saha04a}. For strong
gravitational fields corresponding to small values of the AdS curvature
radius, the contribution from nonzero KK modes along $\Sigma $ is suppressed
by the factor $e^{-2\lambda _{\beta }|z-z_{a}|}$. For the zero KK mode the
components of the brane induced vacuum energy-momentum tensor behave like $%
k_{D}^{D_{1}+1}e^{D_{2}k_{D}y}\exp [(D_{1}+2\nu )k_{D}(y-a)]$ in the region $%
y<a$ and like $k_{D}^{D_{1}+1}e^{D_{2}k_{D}y}\exp [2\nu k_{D}(a-y)]$ in the
region $y>a$. The corresponding quantities integrated over the internal
space contain additional factor $e^{-D_{2}k_{D}y}$ coming from the volume
element and are exponentially small in both regions. For fixed values of the
other parameters, the brane induced VEV in the region $y>a$ vanishes as $%
z_{a}^{2\nu }$ when the brane position tends to the AdS boundary. When the
brane position tends to the AdS horizon, $z_{a}\rightarrow \infty $, for
massive KK modes along $\Sigma $ the VEV of the energy-momentum tensor in
the region $z<z_{a}$ is suppressed by the factor $e^{-2z_{a}\lambda _{\beta
}}$. For the zero mode in the same limit the suppression is power-law with
respect to $z_{a}$.

For the geometry of two branes, the VEV in the region between the branes is
presented as
\begin{equation}
\langle 0|T_{M}^{N}|0\rangle =\langle T_{M}^{N}\rangle
^{(0)}+\sum_{j=a,b}\langle T_{M}^{N}\rangle ^{(j)}+\langle T_{M}^{N}\rangle
^{(ab)},  \label{TMNtwopl1}
\end{equation}%
with separated boundary-free, single branes and interference parts. The
latter is finite everywhere including the points on the branes. The surface
divergences are contained in the single brane parts only. The both single
brane and interference parts separately satisfy the continuity equation and
are traceless for a conformally coupled massless scalar. The possible trace
anomalies are contained in the boundary-free parts. In the limit $%
k_{D}\rightarrow 0$ we derive the corresponding results for two parallel
Robin plates in the bulk $R^{(D_{1},1)}\times \Sigma $. For small values of
the length scale of the internal space corresponding to large KK masses, the
interference part in the VEV of the energy-momentum tensor is suppressed by
the factor $e^{-2\lambda _{\beta }(z_{b}-z_{a})}$. The interference part
vanishes as $z_{a}^{2\nu }$ when the left brane tends to the AdS boundary.
Under the condition $z\ll z_{b}$ an additional suppression factor appears in
the form $(z/z_{b})^{D_{1}}$ for ${}_{D}^{D}$--component and in the form $%
(z/z_{b})^{D_{1}+2\alpha _{1}}$ for the other components, where $\alpha
_{1}=\min (1,\nu )$.

\section{Interaction forces}

\label{sec:Intforce}

Now we turn to the investigations of the vacuum forces acting on the branes.
The corresponding effective pressure $p^{(j)}$ acting on the brane at $%
z=z_{j}$ is determined by ${}_{D}^{D}$--component of the vacuum
energy-momentum tensor evaluated at the point of the brane location: $%
p^{(j)}=-\langle T_{D}^{D}\rangle _{z=z_{j}}$. For the region between two
branes it can be presented as a sum of two terms: $p^{(j)}=p_{1}^{(j)}+p_{{%
\mathrm{(int)}}}^{(j)}$, $j=a,b$. The first term is the pressure for a
single brane at $z=z_{j}$ when the second brane is absent. This term is
divergent due to the surface divergences in the VEVs and needs additional
renormalization. This can be done, for example, by applying the generalized
zeta function technique to the corresponding mode-sum. Below we will be
concentrated on the term $p_{{\mathrm{(int)}}}^{(j)}$. This term is the
additional vacuum pressure induced by the presence of the second brane, and
can be termed as an interaction force. It is determined by the last term on
the right of formulae (\ref{Tikjint}) evaluated at the brane location $%
z=z_{j}$. It is finite for all nonzero interbrane distances and is not
changed by the renormalization procedure. Substituting $z=z_{j}$ into the
second term on the right of formula (\ref{Tikjint}), for the interaction
part of the vacuum effective pressure one finds
\begin{equation}
p_{{\mathrm{(int)}}}^{(j)}=\frac{k_{D}^{D+1}z_{j}^{D}}{(4\pi )^{\frac{D_{1}}{%
2}}\Gamma \left( \frac{D_{1}}{2}\right) }\sum_{\beta }|\psi _{\beta
}(X)|^{2}\int_{\lambda _{\beta }}^{\infty }duu(u^{2}-\lambda _{\beta }^{2})^{%
\frac{D_{1}}{2}-1}\Omega _{j\nu }(uz_{a},uz_{b})F_{\beta }^{(j)}(uz_{j}),
\label{pintj1}
\end{equation}%
where we have introduced the notation
\begin{equation}
F_{\beta }^{(j)}(u)=\left( u^{2}-\nu ^{2}+2m^{2}/k_{D}^{2}\right)
B_{j}^{2}-D(4\zeta -1)A_{j}B_{j}-A_{j}^{2}-2\left( \zeta -1/4\right)
z_{j}^{2}B_{j}^{2}\eta _{\beta }(X).  \label{Fbetj}
\end{equation}%
For small interbrane distances the interaction part dominates the single
brane parts. For a Dirichlet scalar $\Omega _{j\nu }(uz_{a},uz_{b})>0$ and
the vacuum interaction forces are attractive. For a given value of the AdS
energy scale $k_{D}$ and one parameter manifold $\Sigma $ with size $L$, the
vacuum interaction forces (\ref{pintj1}) are functions on the ratios $%
z_{b}/z_{a}$ and $L/z_{a}$. The first ratio is related to the proper
distance between the branes and the second one is the ratio of the size of
the internal space measured by an observer residing on the brane at $y=a$ to
the AdS curvature radius $k_{D}^{-1}$. The quantity $p_{\mathrm{(int)}}^{(j)}
$ determines the force by which the scalar vacuum acts on the brane due to
the modification of the spectrum for the zero-point fluctuations by the
presence of the second brane. As the vacuum properties depend on the
coordinate $y$, there is no a priori reason for the interaction terms to be
equal for the branes $j=a$ and $j=b$, and the corresponding forces in
general are different even in the case of the same Robin coefficients in the
boundary conditions.

Taking the limit $k_{D}\rightarrow 0$ we obtain the result for the
interaction forces between two Robin plates in the bulk $R^{(D_{1}-1,1)}%
\times \Sigma $. In this case, for a homogeneous internal space the
interaction forces are the same even in the case of different Robin
coefficients for separate branes. For the modes along $\Sigma $ with large
KK masses, the interaction forces are exponentially small. In particular,
for sufficiently small length scales of the internal space this is the case
for all nonzero KK modes and the main contribution to the interaction forces
comes from the zero mode. For small interbrane distances, the interaction
forces are repulsive for Dirichlet boundary condition on one brane and
non-Dirichlet boundary condition on the another and are attractive for other
cases. For small interbrane distances the contribution of the interaction
term dominates the single brane parts, and the same is the case for the
total vacuum forces acting on the branes. When the right brane tends to the
AdS horizon, $z_{b}\rightarrow \infty $, the interaction force acting on the
left brane vanishes as $e^{-2\lambda _{\beta }z_{b}}/z_{b}^{D_{1}/2}$ for
the nonzero KK mode and like $z_{b}^{-D_{1}-2\nu }$ for the zero mode. In
the same limit the corresponding force acting on the right brane behaves as $%
z_{b}^{D_{2}+D_{1}/2+1}e^{-2\lambda _{\beta }z_{b}}$ for the nonzero KK mode
and like $z_{b}^{D_{2}-2\nu }$ for the zero mode. In the limit when the left
brane tends to the AdS boundary the contribution of a given KK mode into the
vacuum interaction force vanishes as $z_{a}^{D+2\nu }$ and as $z_{a}^{2\nu }$
for the left and right branes, respectively. For small values of the AdS
curvature radius corresponding to strong gravitational fields, under the
conditions $\lambda _{\beta }z_{a}\gg 1$ and $\lambda _{\beta
}(z_{b}-z_{a})\gg 1$, the contribution to the interaction forces is
suppressed by the factor $e^{-2\lambda _{\beta }(z_{b}-z_{a})}$. For the
zero KK mode, the corresponding interaction forces integrated over the
internal space behave as $k_{D}^{D_{1}+1}\exp [(D_{1}\delta _{j}^{a}+2\nu
)k_{D}(a-b)]$ for the brane at $y=j$ and are exponentially small. In the
model without the internal space the suppression is relatively weaker.

\section{Surface energy-momentum tensor}

\label{sec:SurfEMT}

On manifolds with boundaries the energy-momentum tensor in addition to the
bulk part contains a contribution located on the boundary. For an arbitrary
smooth boundary $\partial M$ with the inward-pointing unit normal vector $%
n^{L}$, the surface part of the energy-momentum tensor for a scalar field is
given by the formula \cite{Saha04c} $T_{MN}^{\mathrm{(s)}}=\delta
(x;\partial M)\tau _{MN}$, where the 'one-sided' delta-function $\delta
(x;\partial M)$ locates this tensor on $\partial M$ and
\begin{equation}
\tau _{MN}=\zeta \varphi ^{2}K_{MN}-(2\zeta -1/2)h_{MN}\varphi n^{L}\nabla
_{L}\varphi .  \label{tausurf}
\end{equation}%
In this formula, $h_{MN}$ is the induced metric on the boundary and $K_{MN}$
is the corresponding extrinsic curvature tensor. From the point of view of
physics on the brane at $y=j$, Eq. (\ref{tausurf}) corresponds to the
gravitational source of the cosmological constant type with the surface
energy density $\varepsilon _{j}^{{\mathrm{(s)}}}=\langle 0|\tau
_{0}^{(j)0}|0\rangle $ (surface energy per unit physical volume on the brane
at $y=j$ or brane tension), stress $p_{j}^{{\mathrm{(s)}}}=-\langle 0|\tau
_{1}^{(j)1}|0\rangle $, and the equation of state $\varepsilon _{j}^{{%
\mathrm{(s)}}}=-p_{j}^{{\mathrm{(s)}}}$. It is noteworthy that this relation
takes place for both subspaces on the brane.

For two-brane geometry the VEV of the surface energy density on the brane at
$y=j$ is presented as the sum $\varepsilon _{j}^{{\mathrm{(s)}}}=\varepsilon
_{1j}^{{\mathrm{(s)}}}+\Delta \varepsilon _{j}^{{\mathrm{(s)}}}$. The first
term on the right is the energy density induced on a single brane when the
second brane is absent. This part is evaluated in \cite{Saha07} by using the
generalized zeta function method. The second term is induced by the presence
of the second brane and is given by the formula
\begin{equation}
\Delta \varepsilon _{j}^{{\mathrm{(s)}}}=\frac{%
2C_{j}n^{(j)}(k_{D}z_{j})^{D}B_{j}^{2}}{(4\pi )^{D_{1}/2}\Gamma \left(
D_{1}/2\right) }\sum_{\beta }|\psi _{\beta }(X)|^{2}\int_{\lambda _{\beta
}}^{\infty }du\,u(u^{2}-\lambda _{\beta }^{2})^{\frac{D_{1}}{2}-1}\Omega
_{j\nu }(uz_{a},uz_{b}),  \label{emt2pl3}
\end{equation}%
with the notation $C_{j}=\zeta -(2\zeta -1/2)\tilde{A}_{j}/(k_{D}\tilde{B}%
_{j})$. As we consider the region $a\leqslant y\leqslant b$, the energy
density $\varepsilon _{j}^{{\mathrm{(s)}}}$ is located on the surface $y=a+0$
for the left brane and on the surface $y=b-0$ for the right brane. The
energy densities on the surfaces $y=a-0$ and $y=b+0$ are the same as for the
corresponding single brane geometry. For an observer living on the brane at $%
y=j$ the corresponding effective $D_{1}$-dimensional cosmological constant
is determined by the relation
\begin{equation}
\Lambda _{D_{1}j}=8\pi M_{D_{1}j}^{2-D_{1}}e^{-D_{2}k_{D}j}\int_{\Sigma
}d^{D_{2}}X\sqrt{\gamma }\,\Delta \varepsilon _{j}^{{\mathrm{(s)}}},
\label{effCC}
\end{equation}%
where $M_{D_{1}j}$ is the $D_{1}$-dimensional effective Planck mass scale
for the same observer. In Ref. \cite{Saha07} it has been shown that for
large distances between the branes the induced surface densities give rise
to an exponentially suppressed cosmological constant on the brane. In the
Randall-Sundrum braneworld model, for the interbrane distances solving the
hierarchy problem between the gravitational and electroweak mass scales, the
cosmological constant generated on the visible brane is of the right order
of magnitude with the value suggested by the cosmological observations.

On background of manifolds with boundaries the total vacuum energy is
splitted into bulk and boundary parts. In the region between two branes the
bulk energy per unit coordinate volume in the $D_{1}$-dimensional subspace
is obtained by the integration of the ${_{0}^{0}}$-component of the volume
energy-momentum tensor over this region: $E^{\text{\textrm{(v)}}}=\int
d^{D_{2}}Xdy\,\sqrt{|g|}\langle 0|T_{0}^{\text{\textrm{(v)}}0}|0\rangle $.
The surface energy per unit coordinate volume in the $D_{1}$-dimensional
subspace, $E^{\text{\textrm{(s)}}}$, is related to the surface densities by
the formula $E^{\text{\textrm{(s)}}}=\sum_{j=a,b}(k_{D}z_{j})^{-D}%
\varepsilon _{j}^{\text{\textrm{(s)}}}$. Now it can be seen that the formal
relation $E=E^{\text{\textrm{(v)}}}+E^{\text{\textrm{(s)}}}$ takes place for
the unrenormalized VEVs, where
\begin{equation}
E=\frac{1}{2}\int \frac{d^{D_{1}-1}{\mathbf{k}}}{(2\pi )^{D_{1}-1}}%
\sum_{\beta }\sum_{n=1}^{\infty }(k^{2}+m_{n}^{2}+\lambda _{\beta
}^{2})^{1/2},\;m_{n}=\gamma _{\nu ,n}/z_{a},  \label{toten1}
\end{equation}%
is the total vacuum energy per unit coordinate volume of the $D_{1}$%
-dimensional subspace, evaluated as the sum of zero-point energies of
elementary oscillators. The latter can be presented in the form $%
E=\sum_{j=a,b}E_{j}+\Delta E$, where $E_{a}$ ($E_{b}$) is the vacuum energy
for the geometry of a single brane at $y=a$ ($y=b$) in the region $%
y\geqslant a$ ($y\leqslant b$), and the interference term is given by the
formula%
\begin{equation}
\Delta E=\sum_{\beta }\int_{\lambda _{\beta }}^{\infty }du\,\frac{%
u(u^{2}-\lambda _{\beta }^{2})^{D_{1}/2-1}}{(4\pi )^{D_{1}/2}\Gamma \left(
D_{1}/2\right) }\ln \left\vert 1-\frac{\bar{I}_{\nu }^{(a)}(uz_{a})\bar{K}%
_{\nu }^{(b)}(uz_{b})}{\bar{K}_{\nu }^{(a)}(uz_{a})\bar{I}_{\nu
}^{(b)}(uz_{b})}\right\vert .  \label{DeltaEhigh}
\end{equation}%
The total vacuum energy within the framework of the Randall-Sundrum
braneworld is evaluated in Refs. \cite{Gold00,Toms00,Flac01b} by the
dimensional regularization method and in Ref. \cite{Garr01} by the zeta
function technique. Refs. \cite{Gold00,Toms00,Garr01} consider the case of a
minimally coupled scalar field in $D=4$, and the case of arbitrary $\zeta $
and $D$ with zero mass terms $c_{a}$ and $c_{b}$ is discussed in Ref. \cite%
{Flac01b}. For the orbifolded version of the model under consideration with $%
D_{1}=4$ and zero mass terms on the branes, the vacuum energy is
investigated in \cite{Flac03b} by using the dimensional regularization. The
zeta function approach in the general case is considered in \cite{Saha07}.

Now let us check that for the separate parts of the vacuum energy the
standard energy balance equation takes places. We denote by $P$ the
perpendicular vacuum stress on the brane integrated over the internal space.
This stress is determined by the vacuum expectation value of the ${}_{D}^{D}$%
-component of the bulk energy-momentum tensor: $P=-\int d^{D_{2}}X\,\sqrt{%
\gamma }\langle 0|T_{D}^{\text{\textrm{(v)}}D}|0\rangle $. In the presence
of the surface energy the energy balance equation is in the form
\begin{equation}
dE=-PdV+\sum_{j=a,b}E_{j}^{\text{\textrm{(s)}}}dS^{(j)},\;E_{j}^{\text{%
\textrm{(s)}}}=\int d^{D_{2}}X\,\sqrt{\gamma }\varepsilon _{j}^{\text{%
\textrm{(s)}}},  \label{enbalance}
\end{equation}%
where $V$ is the $(D+1)$-volume in the bulk and $S^{(j)}$ is the $D$-volume
on the brane $y=j$ per unit coordinate volume in the $D_{1}$-dimensional
subspace:
\begin{equation}
V=\int_{a}^{b}dye^{-Dk_{D}y}\int d^{D_{2}}X\,\sqrt{\gamma }%
,\;S^{(j)}=e^{-Dk_{D}j}\int d^{D_{2}}X\,\sqrt{\gamma },\quad j=a,b.
\label{Sj}
\end{equation}%
It can be explicitly checked that the separate parts in the vacuum energies
and effective pressures on the branes obey the equation (\ref{enbalance}).

\section{Conclusion}

From the point of view of embedding the braneworld model into a more
fundamental theory one may expect that a more complete version of this
scenario must admit the presence of additional extra dimensions compactified
on a manifold $\Sigma $. In the present paper we have considered the local
vacuum effects in the braneworlds with the AdS bulk on a higher dimensional
brane models which combine both the compact and warped geometries. This
problem is also of separate interest as an example with gravitational,
topological, and boundary polarizations of the vacuum, where one-loop
calculations can be performed in closed form. We have investigated the
Wightman function and the bulk and surface Casimir densities for a scalar
field with an arbitrary curvature coupling parameter satisfying Robin
boundary conditions on two parallel branes in $AdS_{D_{1}+1}\times \Sigma $
spacetime. In the region between the branes the KK modes corresponding to
the radial direction are zeros of a combination of the cylinder functions.
The application of the generalized Abel-Plana formula to the corresponding
mode sum allowed us to extract from the VEVs the boundary-free part and to
present the brane induced parts in terms of integrals rapidly convergent in
the coincidence limit of the arguments. We give an application of our
results to the higher dimensional version of the Randall-Sundrum braneworld
with arbitrary mass terms on the branes. For the untwisted scalar the Robin
coefficients are expressed through these mass terms and the curvature
coupling parameter by formulae (\ref{AtildeRS}). For the twisted scalar
Dirichlet boundary conditions are obtained on both branes.

In the model under discussion the hierarchy between the fundamental Planck
scale and the effective Planck scale in the brane universe is generated by
the combination of redshift and large volume effects. For large interbrane
separations the corresponding effective Newton's constant on the brane at $%
y=b$ is exponentially small. This mechanism also allows obtaining a
naturally small cosmological constant generated by the vacuum quantum
fluctuations of a bulk scalar. In \cite{Saha07} we have considered two
classes of models with the compactification scale on the visible brane close
to the fundamental Planck scale. For the first one the higher dimensional
Planck mass and the AdS inverse radius are of the same order and in the
second one a separation between these scales is assumed. In both cases the
corresponding interbrane distances generating the hierarchy between the
electroweak and Planck scales are smaller than those for the model without
an internal space and the required suppression of the cosmological constant
is obtained without fine tuning.

\section*{Acknowledgments}

The work was supported by the Armenian Ministry of Education and Science,
Grant No. 119.

\end{document}